\documentstyle[aps,twocolumn]{revtex}

\begin{document}
\title{Metallic and insulating behaviour of the two-dimensional electron gas on a
vicinal surface of Si MOSFETs}
\author{S. S. Safonov$^1$, S. H. Roshko$^1$, A. K. Savchenko$^1$, A. G. Pogosov$%
^{1,2}$, and Z. D. Kvon$^2$}
\address{$^1$ School of Physics, University of Exeter, Stocker Road,\\
Exeter, EX4 4QL, UK\\
$^2$ Institute of Semiconductor Physics, \\
Siberian Branch Russian Academy of Sciences, 630090\\
Novosibirsk, Russia}
\maketitle

\begin{abstract}
The resistance $R$ of the 2DEG on the vicinal Si surface shows an unusual
behaviour, which is very different from that in the $\left( 100\right) $ Si
MOSFET where an unconventional metal to insulator transition has been
reported. The crossover from the insulator with $dR/dT<0$ to the metal with $%
dR/dT>0$ occurs at a low resistance of $R_{\Box }^c$ $\sim 0.04\times h/e^2$
. At the low-temperature transition, which we attribute to the existence of
a narrow impurity band at the interface, a distinct hysteresis in the
resistance is detected. At higher temperatures, another change in the sign
of $dR/dT$ is seen and related to the crossover from the degenerate to
non-degenerate 2DEG.
\end{abstract}


The problem of the metal-to-insulator transition (MIT) in two-dimensional
systems has been attracting much attention after the observation that in the
two-dimensional gas in (100) Si MOSFET there is a change in the sign of the
temperature dependence from $dR/dT>0$ (metal) to $dR/dT<0$ (insulator) with
varying concentration \cite{Kravchenko}. The transition occurs at a
resistance of $R_{\Box }\sim h/e^2$ and was then confirmed to exist in other 
{\em n}-(100) Si \cite{Popovic} as well as in Si/SiGe and {\em p}-GaAs
structures \cite{Coleridge,Simmons}. The metallic-like behaviour is in
obvious contradiction with the 2D scaling theory of electron localization 
\cite{Gang4} which allows only insulating behaviour. The specifics of the
unusual metal is in the fact that it has been seen in high mobility
structures and at low concentration of carriers which have large effective
mass, so that the Coulomb interaction could play a significant role. There
have been many different explanations suggested \cite{All-theory} although
this effect still remains an unresolved problem. Recently, several models
have appeared where the unusual metallic behaviour is explained by
conventional, though non-trivial, electron transport at low temperatures: i)
classical conduction with scattering by an impurity band in the oxide \cite
{Altshuler!MaslovPRL}, ii) temperature dependent screening of impurity
scattering and crossover from the degenerate to non-degenerate state \cite
{DasSarmaPRL}, iii) two-band conduction in {\em p}-GaAs \cite{Sivan}.

In a broad temperature range from 50 mK to 70 K, we have performed an
investigation of the MIT on the 2DEG in high mobility Si MOSFETs with {\em %
another }orientation of Si surface - the {\em vicinal }surface which is cut
at a small angle to the plane $(100)$. Such types of structure have been
studied previously in the context of superlattice effects \cite
{Ando!Fowler!Stern,Volkov,Kvon} which were seen at higher electron
concentrations than used in this work. We expected that the difference in
the surface and impurity states at the interface would affect the
manifestation of the MIT. Indeed, our results show that the MIT in the 2DEG
of Si MOSFETs is {\em not universal} and has a different manifestation in
the vicinal samples. We have observed $two$ crossovers in $R(T)$, at low and
high temperatures, which are explained in terms of the models \cite
{Altshuler!MaslovPRL} and \cite{DasSarmaPRL} of the temperature dependent
impurity scattering. The low-temperature transition has been seen at a small
critical resistance where one can neglect quantum corrections to the
conductivity. We have observed a strong hysteresis at this transition, which
clearly indicates that it originates from a narrow impurity band (IB). We
also report an unusual low-temperature reentrant MIT, which does not exist
in (100) Si structures we made by the same technology for a comparative
study.

The vicinal samples are high mobility {\it n}-channel MOSFETs fabricated on
a surface which is tilted from the $\left( 100\right) $ surface around the $%
\left[ 011\right] $ direction by an angle of $9.5^{\circ }$. The samples
have a peak mobility of $2\times 10^4$ $cm^2/Vs$ at $T=4.2$ K. The `normal'
samples are grown on the $\left( 100\right) $ Si and have maximum mobility
around $1.5\times 10^4$ $cm^2/Vs$. The oxide thickness in both types of
structure is $120$ $nm$. The samples have a Hall bar geometry with length $%
1200$ $\mu m$ and width $400$ $\mu m$. Their resistance has been measured in
the temperature range $0.05-70$ K by a four-terminal {\it ac} method with
frequency $\leq 10$ Hz and current $2\leq I_{ac}\leq 10$ $nA$. The electron
concentration has been determined by the Shubnikov-de Haas and capacitance
measurements, and has been varied in the range $2\times 10^{11}-1.4\times
10^{12}$ $cm^{-2}$.

Fig. 1 shows the resistance as a function of the gate voltage $V_g$ for a
vicinal sample Si-4.1 in the temperature range below 1 K. A change in the
sign of $dR/dT$ is clearly seen near $R_{\Box }\sim 1$ kOhm $\sim 0.04\times
h/e^2$, with metallic behaviour at larger $V_g$. When the gate voltage,
controlling the concentration, is slowly swept (with rate $2$ $V$/hour) in
the two opposite directions, two distinct groups of curves are detected. The
hysteresis loop disappears above 4 K and seems to be most pronounced near
the crossover region. To quantify this observation, we have performed an
experiment where a particular $V_g$ is approached from opposite directions:
from $V_g^{(1)}=0.5$ V and $V_g^{(2)}=9$ V. After a brief transient time
when the equilibrium is established, the difference between the two
resistances $\Delta R=\mid R^{(1)}-R^{(2)}\mid $ is not changing for many
hours. This value is shown in Fig. 1, inset, with a clear peak at $%
V_g\approx 2.2$ V - exactly in between the two crossover points. The
Shubnikov-de Haas measurements performed in each case have shown that, for a
particular $V_g$, the electron concentration is independent of the direction
of the sweep - that is, it is the difference in the mobility which gives
rise to $\Delta R$.

Noticeably, the crossover point in the same sample does not have a universal
nature: the two transition points do not coincide either in their resistance
or concentration. On the other hand, the value of the mobility is
practically the same at the transition points, which indicates that it is
the mobility which governs the transition. Hence we suggest that the peak in 
$\Delta R$ occurs when a narrow ($W<0.5$ meV) impurity band at the interface
comes to the Fermi level of the 2DEG and changes the character of the
electron scattering. A natural suggestion for the origin of the hysteresis
is a slow (at low temperatures) electron exchange between the impurity band
and the 2DEG separated by a barrier. With increasing $V_g$ and rising of the
Fermi level, the IB\ gets charged by electrons from the 2DEG. Some of the
electrons will still remain in the IB\ when the gate voltage is decreased
back to lower values, until the Fermi level is below the IB and it releases
all its negative charge (this is why $\Delta R$ is small both at high and
low $V_g$s ).

{It is worth mentioning that the presence of the IB was detected earlier in
(100) Si MOSFETs \cite{Ando!Fowler!Stern}, although this was done in the
hopping regime}, {where it gave rise to an increased density of localized
states and was easily detected as a peak in the conductance }$G(V_g)$. Here
its effect is seen on electron scattering in the metallic regime: as a
crossover point in $R(T)$ and as a peak in the hysteresis.

We suggest that the character of the IB\ is similar to that considered in {%
\cite{Altshuler!MaslovPRL}: it scatters electrons when it is positively
charged and the scattering decreases when more electrons are added to it. In
that model, when the Fermi level is above and close to the impurity band,
the IB does not contribute to scattering at }${T=0}$ K{. With increasing
temperature, the IB becomes positively charged and the resistance of the
2DEG increases, }$dR/dT>0$. To explain the transition to the insulating
behaviour with decreasing electron concentration, it was assumed that the
electron localization takes over{\ }at large enough resistance of the order
of $R_{\Box }\sim $10 kOhm \cite{Altshuler!Maslov!Pudalov}. In our case,
however, the transition occurs at much lower resistance where one can
neglect electron localization. At the same time, we think that for a {\em %
narrow} {\em band} the crossover in the sign of $dR/dT$ should occur when
the Fermi level $F$ is close to its centre. When the Fermi level moves down
to the lower part of the IB, the mobility $\mu $ will increase with
increasing temperature, as the positive charge of the IB decreases as $\mu
^{-1}(T)\propto N^{+}\propto 1/\left( 1+\exp \frac{F-E_i}{k_BT}\right) $,
where the IB is assumed to be at the level $E_i$, Fig. 2, inset. However, in
such a simple model one should not expect a decrease in the resistance by
more than a factor of two.

Fig. 2 shows the temperature dependence of the resistance $R^{(2)}$ near the
transition at $n_c\simeq 4.18\times 10^{11}$ $cm^{-2}$, in the range T=$50$
mK $-4$ K, presented as $\mu ^{-1}$ to illustrate this simple model of IB
scattering. To calculate the IB\ contribution, we subtracted the background $%
\mu ^{-1}$ at $V_g=2.4$ V when the IB is full. The value $F-E_i$ is
calculated from known electron concentration and hence the Fermi energy $E_F$%
. The IB has been taken as having a constant density of states with width $%
W\simeq 0.08$ meV used as an adjustable parameter. It is interesting to note
that in order to get a satisfactory agreement, we have to shift gradually
the level $E_i$ up when the IB gets more than half filled. Also, the width
of the IB appears to be smaller for the set $R^{(2)}(T)$ (by a factor of
two), that is, when the IB is more filled with electrons. We expect this to
be a reflection of the Coulomb interaction of the states in the IB (to be
discussed elsewhere \cite{AKS}). It is important to emphasize that the
`insulating' behaviour presented in Fig. 2 for the resistance range $%
R=0.9-1.2$ kOhm is in fact a property of a metallic 2DEG with a well defined
Fermi surface. A direct proof of this is obtained by measuring Shubnikov-de
Haas oscillations at $V_g=2.0$ V $-$ for a concentration below the
transition.

Fig. 3 shows the temperature dependence in the whole temperature range. It
is seen that the discussed low-temperature MIT at $R\sim 1$ kOhm exists only
in a narrow range of temperatures and that, in general, $R(T)$ has
complicated non-monotonic character. At $V_g<2$ V, the 2DEG shows an
insulator which cannot be explained by the simple IB model and cannot occur
due to the electron localization of a quantum nature because of the low
sample resistance. This could be a result of a percolation-type
localization. In Fig. 3, inset, the variation of the slope of the
temperature dependence with decreasing $V_g$ is shown for this insulator,
using an exponential fit of $R(T)$ in the range $T\sim 1.5-7$ K. The
activation energy $\Delta =E_c-F$ is seen to increase linearly with
decreasing Fermi level (calculated from the capacitance consideration). The
activation energy extrapolates to zero at the mobility edge $E_c$
corresponding to $V_g=1.9$ V - the value which is close to the lowest $V_g$
where Shubnikov- de Haas oscillations were seen.

Let us discuss the behaviour of $R\left( T\right) $ in Fig. 3 with
increasing temperature. Localized electrons become delocalized at $k_BT\geq
\Delta $, and after a dip in $R(T)$ around $T\sim 4$ K there is a change in
the sign of the temperature dependence and the transition to metallic
behaviour. At higher temperatures, another crossover is now seen at $R_{\Box
}^C$ $\simeq $ 3 kOhm. Near this transition, there is a non-monotonic $%
R\left( T\right) $ with a gradual change from $dR/dT>0$ to $dR/dT<0$ with
increasing $T$. The phonon scattering can be neglected in this regime as it
only becomes important at $T>100$ K \cite{Ando!Fowler!Stern}. We note that
in the temperature range $T>$ 4 K the system experiences a transition from
degenerate (quantum) to nondegenerate (classical) state (the Fermi
temperature $T_F$ varies from $90.5$ K for the bottom curve to $15$ K for
the top one in Fig. 3). The variation of $T_F$ with concentration
corresponds to the position of the hump in Fig. 3. All main features of the
model {\cite{DasSarmaPRL} for the t}emperature dependent ionized impurity
scattering can be seen in this regime. The metallic behaviour at $T<T_F$ is
explained in {\cite{DasSarmaPRL} }by the temperature dependence of the
screening function. This produces a linear rise in the resistivity with
increasing temperature, $\rho \left( T\right) \propto \rho \left( T=0\right)
+A\left( T/T_F\right) $, which agrees with experiment where $\rho \left(
T\right) \propto \left( T/T_F\right) ^{1.1}$ for the bottom curve in Fig. 3.
When $T>T_F$, the temperature dependence is expected to become $\rho \left(
T\right) \propto A\left( T/T_F\right) ^{-1}$ due to ionized impurity
scattering in the non-degenerate case which is in a qualitative agreement
with upper curves in Fig. 3.

The further support of the classical to quantum transition at high
temperatures is obtained from measurements of the perpendicular
magnetoresistance. When $T>T_F$, we observed a small, $\Delta R/R\sim $1-2
\%, positive magnetoresistance. The magnetoresistance decreases with either
increasing concentration or decreasing temperature, i.e. when the system is
driven towards the degenerate state. This qualitatively agrees with the
classical behaviour of degenerate semiconductors \cite{Kireev}.

We have performed a comparative study of a normal sample (100) Si \cite{AKS}%
. At high temperatures it also shows a crossover in $R(T)$ but at a higher
critical resistance, $R_c\sim 15$ kOhm, and lower concentration, $n_c\simeq
2.8\times 10^{11}$ $cm^{-2}$, than in the vicinal sample. The position of
the resistance hump near the transition is shifted from $T_F\sim 20$ K in
the vicinal sample to $T_F\sim 10$ K in the normal sample. This supports the
applicability of the model \cite{DasSarmaPRL} for the high-temperature
transition in both samples. Comparing our normal sample with other (100) Si
samples, one can see a similarity in the shape of $R(T)$ near the
transition, apart from the fact that the crossover in $R(T)$ is shifted from 
$T\sim $ 2 K in \cite{Kravchenko} to $T\sim 10$ K in our normal sample. If
it is assumed that the transition in \cite{Kravchenko} can also be explained
in terms of \cite{DasSarmaPRL}, the difference in the concentrations ($%
n_c\sim 1\times 10^{11}$ $cm^{-2}$ \cite{Kravchenko}) could account for the
shift of the transition in the temperature scale.

At temperatures below 1 K, the normal sample shows a similar crossover
behaviour as the vicinal ones, although no hysteresis has been observed,
which does not allow us to link directly the low-temperature crossover
around $R\sim 1$ kOhm in the normal sample to an IB. In the metallic regime
below $R\sim 1$ kOhm, we have observed a striking difference between the
vicinal and normal samples: in the vicinal samples there exists another
crossover point at $R\simeq 0.33$ kOhm and $n\sim 8\times 10^{11}$ $cm^{-2}$%
, Fig. 4a. Re-appearance of the insulating state with increasing carrier
concentration has been previously reported in (100) Si-MOS structures \cite
{Pudalov} and p-GaAs/GaAlAs \cite{Hamilton} where it was attributed to weak
electron localization. The effect we have observed on the vicinal sample is
quite different from that in \cite{Pudalov,Hamilton}. Firstly, it shows a
significantly stronger (by a factor of ten)\ insulating $R\left( T\right) $
which cannot be explained by weak localization. Secondly, the new transition
is accompanied by a hump in $R(V_g)$ at base temperature, Fig. 4b. This
feature, which we have seen in several vicinal samples, is possibly a
manifestation of a gap in the energy spectrum which is only detected below 1
K. It is tempting to link this effect with a superlattice minigap, however
this is usually seen at much higher concentrations $\sim 2.5\times 10^{12}$ $%
cm^{-2}$ \cite{Ando!Fowler!Stern,Volkov}. Also, the reentrant insulator
cannot be explained by occupation of the second subband \cite{Sivan} as we
have not been able to identify its presence in the Shubnikov-de Haas
oscillations and expect it to appear at higher electron concentrations \cite
{Ando!Fowler!Stern}.

In conclusion, we have observed several unusual features of the
metal-to-insulator transition of the 2DEG on a vicinal Si surface and have
been able to explain most of them by classical electron conduction with
temperature dependent impurity scattering.

We are grateful to B. L. Altshuler and D. L. Maslov for stimulating
discussions, EPSRC and ORS award fund for financial support. We also thank
Y. Y. Proskuryakov for participating in discussions and helping with
experiment.

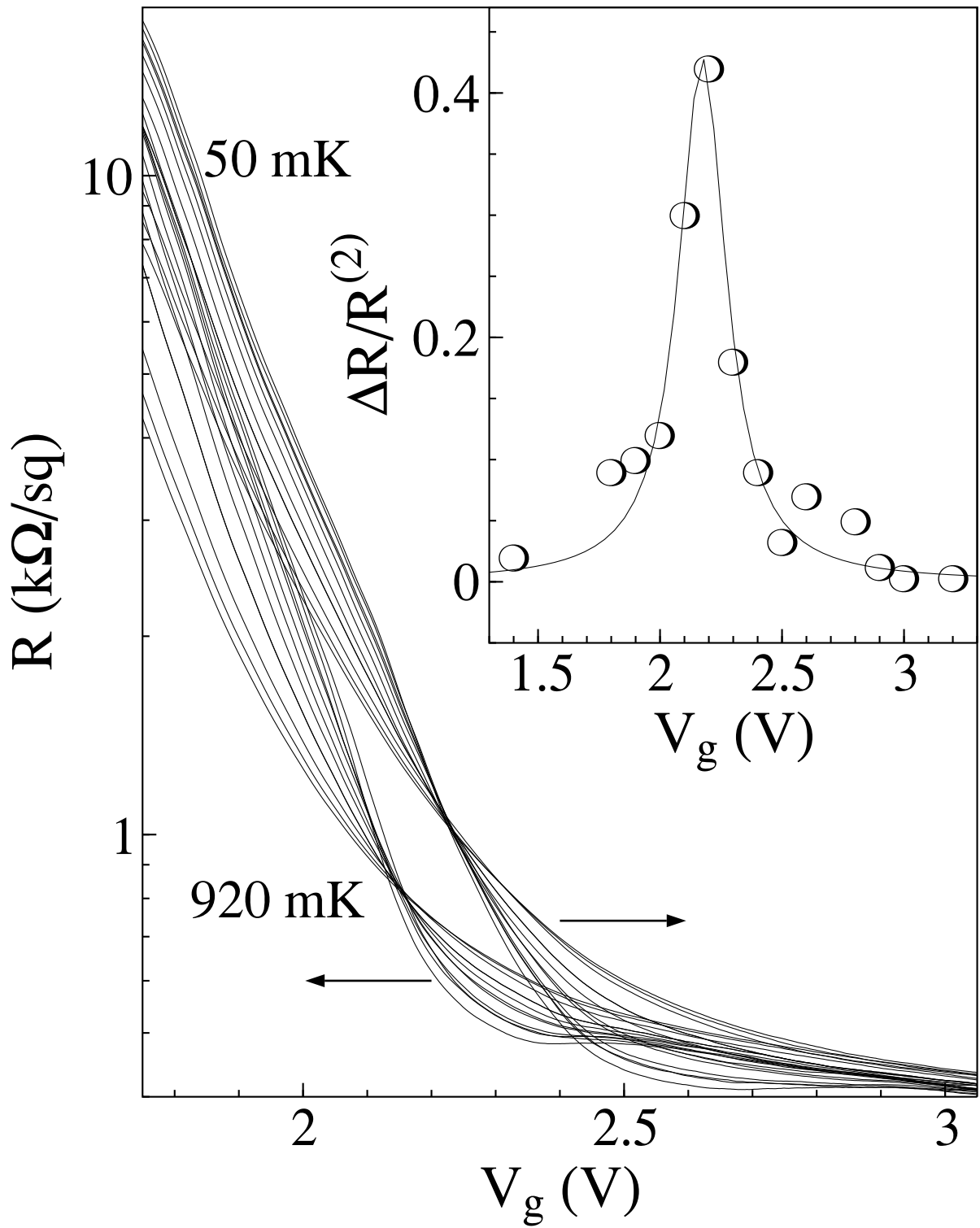
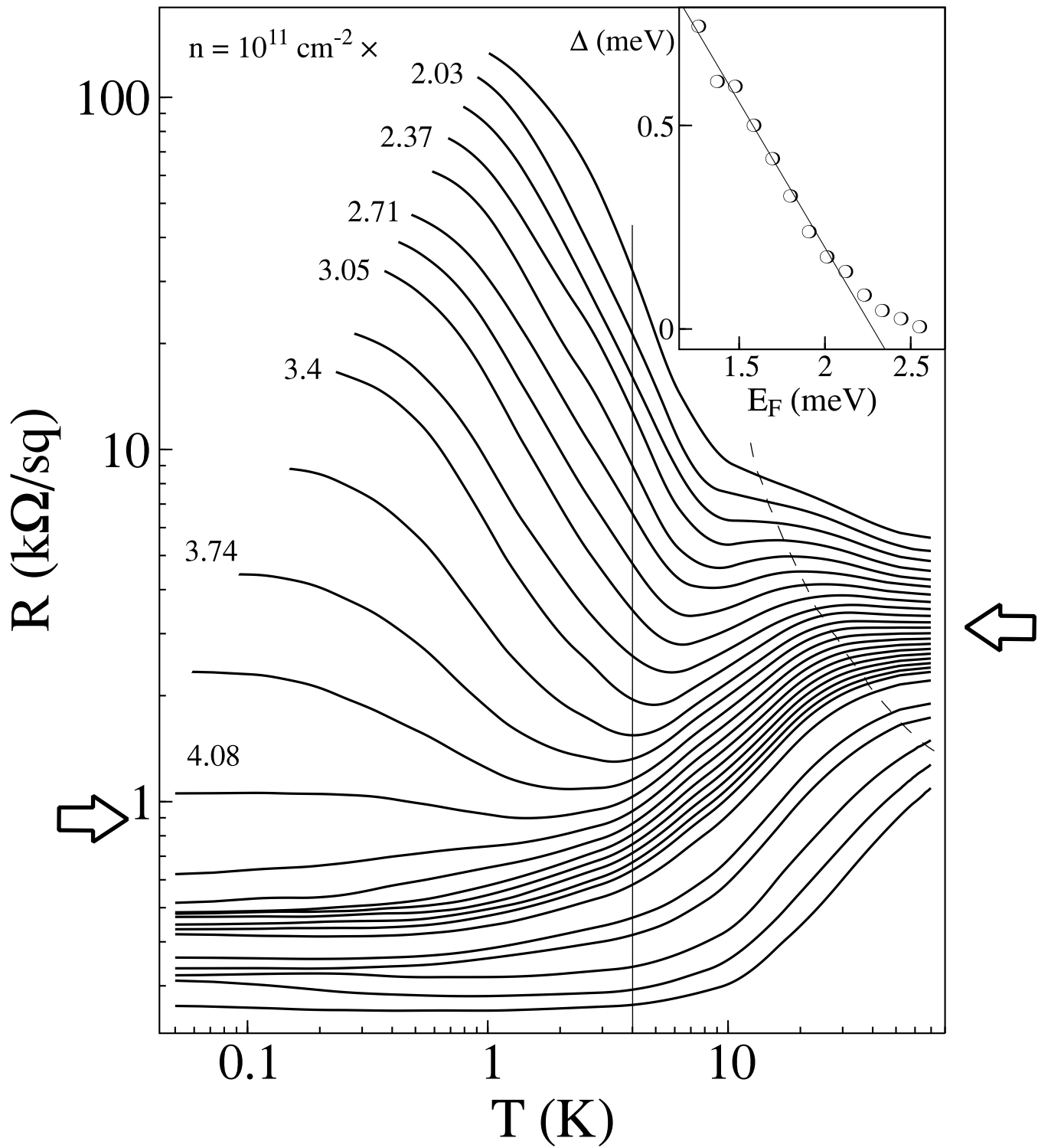
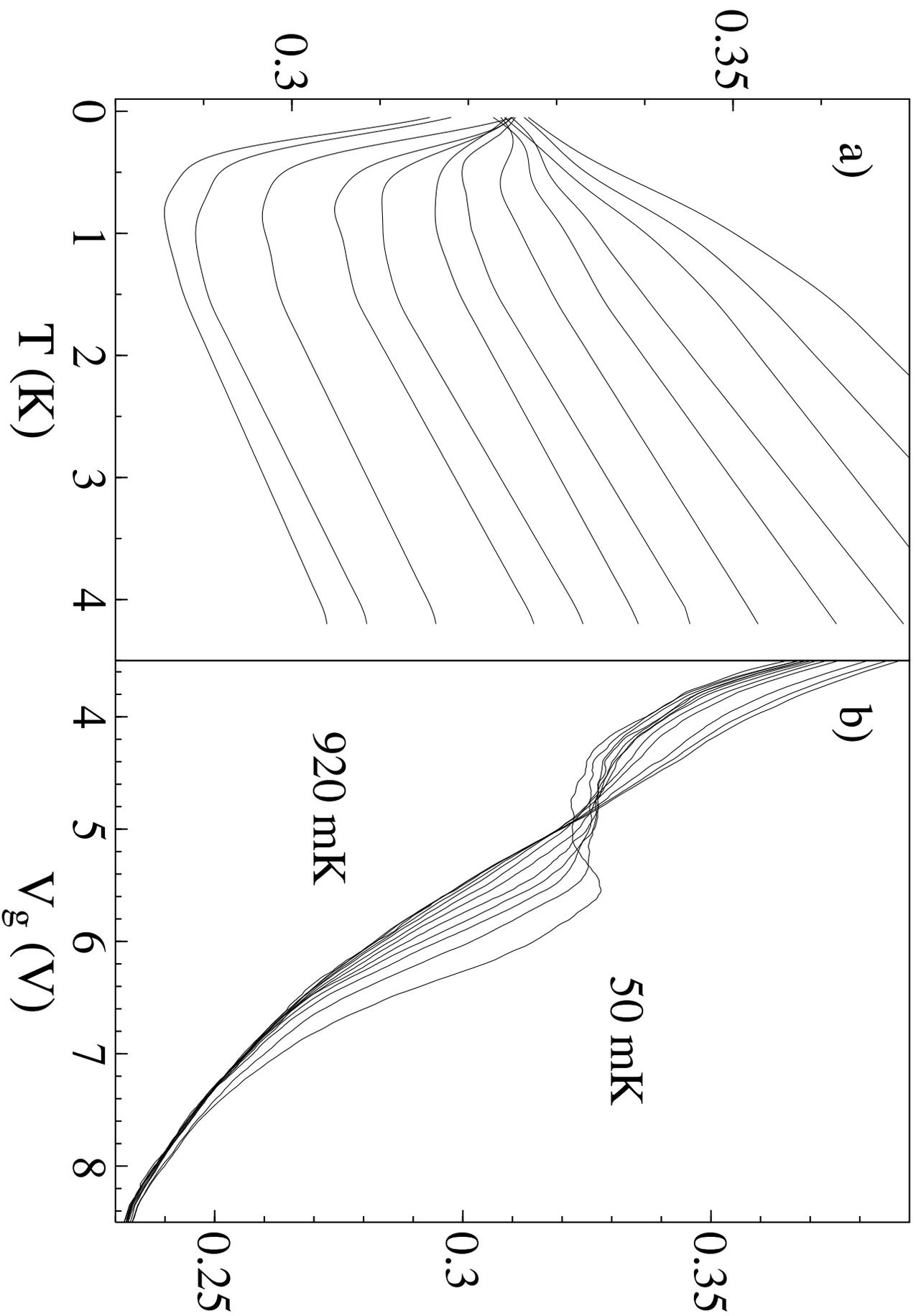
\end{document}